 \documentclass[aps,pra,twocolumn,groupedaddress,amsmath,amssymb,showpacs]{revtex4}
\usepackage{graphicx}
\usepackage{amsmath}
\usepackage{dcolumn}
\usepackage{bm}

\newcommand{\bx}{\mathbf{x}}

\newcommand{\bq}{\mathbf{q}}

\newcommand{\im}{\mathrm{i}}

\newcommand{\ra}{\rangle}
\newcommand{\la}{\langle}

\begin{document}
\title{Unravelling Two-Photon High-Dimensional  Entanglement}
\author{A. Aiello}
\author{S. S. R. Oemrawsingh}
\author{E. R. Eliel}
\author{J.P. Woerdman}
\affiliation{Huygens Laboratory, Leiden University\\
P.O.\ Box 9504, 2300 RA Leiden, The Netherlands}
\begin{abstract}
We propose an interferometric method to investigate the
non-locality of high-dimensional two-photon orbital angular
momentum states generated by spontaneous parametric down
conversion. We incorporate two half-integer spiral phase plates
and a variable-reflectivity output beam splitter into a
Mach-Zehnder interferometer to build an orbital angular momentum
analyzer. This setup enables testing the non-locality of
high-dimensional two-photon states
 by repeated use of the Clauser-Horne inequality.
\end{abstract}
 \maketitle
%
%
%
%
%
%
%
%
\section{Introduction}
%
Entangled qubits play a key role in many applications of quantum
information \cite{NielsenBook} and quantum cryptography
\cite{Gisin02}. An example of a qubit is the polarization state of
a photon. More generally, a qudit is a quantum system whose state
lies in a $d$-dimensional Hilbert space.  The higher
dimensionality implies a greater potential for applications in
quantum information processing and this explains the continuously
growing interest in methods for creating entangled qudits.

Among these methods, spontaneous parametric down-conversion (SPDC)
appears to be the most reliable one for creating entangled photon
pairs \cite{Kwiat95}.  Recently, several techniques have been used
to create entangled qudits from down-converted photons. For
example, conservation of orbital angular momentum (OAM) in SPDC,
has been used to create  entangled states with $d=3$
\cite{Mair01a,Vaziri02a}, and a time binning method was employed
to realize states with $d=11$ \cite{Gisin02b}. Recently, spatial
degrees of freedom  in SPDC \cite{Law04} have been exploited to
demonstrate  entanglement for the cases $d=4,8$ \cite{Padua} and
$d=6$ \cite{Boyd05a}.

It is well known that  \emph{useful} high-dimensional entanglement
can be witnessed by violation of Bell-type inequalities
\cite{Acin03a} which also furnish a test of nonlocality for a
quantum system. However, tests of $d$-dimensional inequalities for
bipartite quantum systems, require  the use of at least $2d$
detectors which becomes exceedingly difficult (if not impossible)
for large $d$.


In a previous paper \cite{Sumant04b} we proposed an experiment to
show the entanglement of high-dimensional two-photon OAM states,
with \emph{two} detectors only.  This scheme indeed allows to
verify the existence of high-dimensional non-separability, as
demonstrated by our subsequent experimental results
\cite{sumant05c}. In Ref. \cite{Sumant04b} we went on to use a
$2$-dimensional Bell inequality to check the non-locality of our
OAM-entangled photons. In the meantime we have realized that this
implicitly assumes dichotomic variables, a condition that was not
fulfilled by the scheme proposed in \cite{Sumant04b}.
In the present paper, we propose an experimental scheme to
explicitly test the non-locality (namely, the \emph{useful}
entanglement) of very-high-dimensional two-photon OAM states ($d
\sim \infty$), by using just $4$ detectors. The advantages of our
method with respect to those using $2d$ detectors are obvious for
$d>2$. Additionally, we stress that the scheme we propose is
designed to realize dichotomic observables.
The idea is first to project the infinite-dimensional two-photon
state onto several {different} four-dimensional subspaces (in
order to select different four-dimensional two-photon states), and
then to apply the Clauser-Horne (CH) inequality \cite{Clauser74a}
to  each selected state.
It is not obvious \emph{a priori} whether such a  scheme will work
or not.
In fact several legitimate questions can be raised: (i) Does this
dimensional reduction spoil the entanglement of the two-photon
state? (ii) Do selected four-dimensional states maximally violate
the CHSH inequality? (iii) Are  distinct four-dimensional
subspaces equivalent?
In the rest of this paper we will address these questions.
\section{The proposed experiment}
Let us describe the scheme of our proposed  experiment (Fig. 1). A
thin nonlinear crystal yields OAM-entangled  photon pairs, and the
two photons (say $a$ and $b$) are fed into two balanced
Mach-Zehnder interferometers. Each Mach-Zehnder $\mathrm{MZ}_x, \;
(x=a,b)$ is made of a $50/50$ input beam splitter and a
variable-reflectivity output beam splitter, indicated in Fig. 1
with BS and $\mathrm{VBS}_x$, respectively. We denote with $t_x$
and $r_x$ the transmission and reflection coefficients of each
$\mathrm{VBS}_x$ and assume
\begin{subequations}
\begin{eqnarray}
  t_x & = & \cos \theta_x, \label{eq10a}\\
  r_x & = &\im \sin \theta_x \label{eq10b},
\end{eqnarray}
\end{subequations}
where $x = a,b$ and $\theta_x \in [0,2 \pi)$.
Such a VBS can be easily realized, for example,  by exploiting the
polarization degrees of freedom of the SPDC photons. Type I
crystals emit photon pairs with a well defined linear
polarization. Then, the combination of an half-wave plate before
the Mach-Zehnder and a polarizing beam splitter as output BS of
the same interferometer, realizes the desired VBS. Another
possibility
 is to use use a Fabry-P\'{e}rot \'{e}talon whose mirror
separation can be varied, to realize a so-called ``Lorentzian beam
splitter'' \cite{Jeffers93a}, which works as a VBS.

In each channel $i, \, (i=1,2)$ of the interferometer
$\mathrm{MZ}_a$ ($\mathrm{MZ}_b$), there is a spiral phase plate
SPP (complementary spiral phase plate: CSPP), oriented at
$\alpha_i$ ($\beta_i$). In the following we shall restrict our
attention to the case $\chi_2 = \chi_1 + \pi, \; (\chi = \alpha,
\beta)$. The output channel ``$1$'' of the interferometer
$\mathrm{MZ}_x$  is coupled to a single-mode fiber
$\mathrm{F}_{x1}$ which sustains the Laguerre-Gaussian mode
$\mathrm{LG}_0^0$ with waist $w_0$. The output ports of the two
fibers $\mathrm{F}_{a1}$  and $\mathrm{F}_{b2}$ are coupled with
two detectors $\mathrm{D}_{a1}$ and $\mathrm{D}_{b1}$,
respectively, which measure the twin-photon coincidence rate. The
experimental scheme also comprises two pairs of imaging systems
(not shown in Fig. 1), which image the twin photons from the the
crystal to the the SPPs, and from the SPPs to the input port of
the fibers.
\begin{figure}[!htl]
\includegraphics[angle=0,width=8truecm]{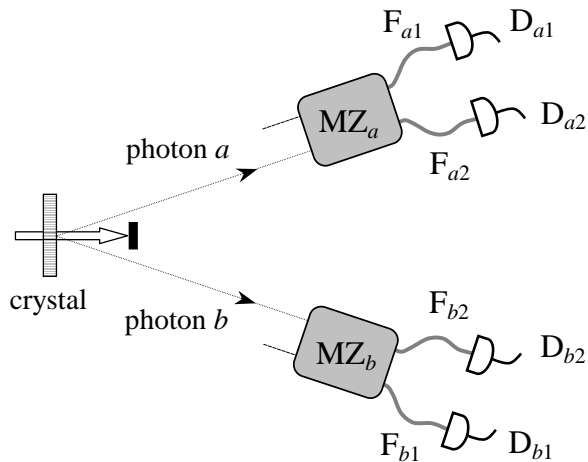}
\caption{\label{fig:1} Schematic of the proposed experimental
setup. The  boxes $\mathrm{MZ}_a$ and $\mathrm{MZ}_b$
 represent the Mach-Zehnder interferometers in the path
of the photon $a$ and $b$, respectively. The thick grey lines
$\mathrm{F}_{xi}$($x=a,b; \; i =1,2$), represent the single-mode
optical fibers. Each of them is coupled with a detector
$\mathrm{D}_{xi}$. Other details are given in the text.}
\end{figure}
\begin{figure}[!hbr]
\includegraphics[angle=0,width=8.5truecm]{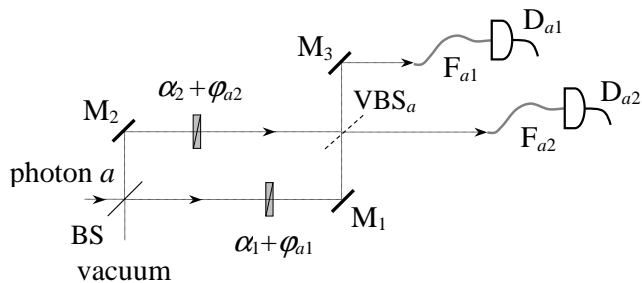}
\caption{\label{fig:2} Detailed scheme of the OAM analyzer in the
path of the photon $a$. BS denotes a $50/50$ beam splitter,
$\mathrm{VBS}_a$ a variable-reflectivity beam splitter, $\alpha_1$
and $\alpha_2$ represent the two SPPs, and $\mathrm{M}_1,
\mathrm{M}_2,\mathrm{M}_3$ represent three ordinary mirrors.   The
role of $\mathrm{M}_3$ is to maintain the input wave function
spatially invariant after an even number of reflections.  Two
additional azimuthal-independent phases $\varphi_{a1}$ and
$\varphi_{a2}$ are accounted for.}
\end{figure}
\begin{figure}[!hbr]
\includegraphics[angle=0,width=6truecm]{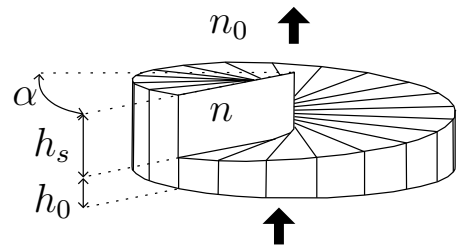}
\caption{\label{fig:3} Schematic drawing of a spiral phase plate
(SPP) with a step index ($=$ phase shift per unit angle)
$\mathcal{L} = h_s(n-n_0)/\lambda$, where $h_s$ is the step
height, $n$ and $n_0$ are the refractive indices of the SPP and
the surrounding medium, respectively, and $\lambda$ is the
wavelength of the incident light. In this letter we assume
$\mathcal{L} = \ell + 1/2, \, \{\ell = 1,2,\ldots \}$. The
orientation angle $\alpha$ is indicated.}
\end{figure}
\subsection{The spiral phase plates}
A spiral phase plate, shown in Fig. 3,  is a transparent
dielectric plate with an edge dislocation that can be freely
rotated around the plate axis \cite{Beije94a}. Let $z$ be the axis
of the plate and $\chi$ the rotation angle. When a light beam with
transverse profile $\psi(\bx)$ crosses such a SPP it acquires an
azimuthal-dependent phase $\exp(\im f(\chi, \phi))$
\begin{equation}\label{eq20}
\psi(\bx) \rightarrow e^{\im f(\chi, \phi)} \psi(\bx),
\end{equation}
where
\begin{equation}\label{eq30}
e^{\im f(\chi, \phi)} = e^{\im \mathcal{L}(\phi - \chi) }\left[
e^{\im 2 \pi  \mathcal{L}} \Theta(\chi - \phi) + \Theta(\phi -
\chi) \right].
\end{equation}
Here $\bx$ is the two-dimensional position vector $\bx = (x,y)$ in
the transverse plane $z=\mathrm{const.}$, $\phi=\phi(\bx)$ is the
azimuthal angle, and $\mathcal{L} \in \mathbb{R} $ is the phase
shift per unit angle. In addition, with $\Theta(X)$ we denoted the
Heaviside function which is equal to $1$ for $X>0$ and  to zero
otherwise.
Let  $\hat{S}(\chi)$ be the quantum mechanical operator
representing the action of a SPP on the arbitrary single-photon
state $|\psi\ra$, and let $| \bx \ra$ denotes the
position-representation state of a quasi-monochromatic photon with
a given polarization (position state, for short)
\begin{equation}\label{eq35}
| \bx \ra =  \frac{1}{2 \pi} \int \mathrm{d}^2 \bq \, e^{-\im \bq
\cdot \bx} \hat{a}^\dagger(\bq)| 0\ra,
\end{equation}
where $\bq = (q_x, q_y)$ is the transverse photon momentum and $[
\hat{a}(\bq), \hat{a}^\dagger(\bq')] = \delta^{(2)}(\bq - \bq')$.
It is easy to see that the position states $\{| \bx \ra  \}$ are
orthogonal and form a complete basis in the single-photon Hilbert
space
\begin{subequations}
\begin{eqnarray}
 \la \bx | \bx' \ra & = & \delta^{(2)}(\bx - \bx'), \label{eq37a}\\
  \hat{I} & = & \int \mathrm{d}^2 \bx \, | \bx \ra \la \bx |.\label{eq37b}
\end{eqnarray}
\end{subequations}
The quantum operator $\hat{S}(\chi)$ can be determined in analogy
with the classical case, by imposing
\begin{equation}\label{eq40}
\la \bx | \hat{S}(\chi) |\psi \ra = e^{\im f(\chi, \phi)}
\psi(\bx),
\end{equation}
and assuming $ \psi(\bx) =\la \bx | \psi \ra $. Then Eq.
(\ref{eq40}) can be rewritten as
\begin{equation}\label{eq50}
\la \bx | \hat{S}(\chi) |\psi \ra = e^{\im f(\chi, \phi)} \la \bx
| \psi \ra ,
\end{equation}
which implies, together with the arbitrariness of $| \psi \ra$,
\begin{equation}\label{eq60}
\la \bx | \hat{S}(\chi) = e^{\im f(\chi, \phi)} \la \bx |.
\end{equation}
This equation shows that the SPP operator $\hat{S}(\chi)$ is
diagonal in the coordinate basis kets and it is unitary since its
eigenvalues $\exp({\im f(\chi, \phi)})$ have modulus $1$. If we
multiply both sides of Eq. (\ref{eq60}) by $| \bx '\ra$,  we
obtain
\begin{equation}\label{eq70}
\begin{array}{rcl}
\la \bx | \hat{S}(\chi) | \bx' \ra & = & e^{\im f(\chi, \phi)} \la \bx | \bx' \ra \\\\
& = & e^{\im f(\chi, \phi')} \la \bx | \bx' \ra,
\end{array}
\end{equation}
where  the second line of Eq. (\ref{eq70}) immediately follows
from the orthogonality of the position states. From Eq.
(\ref{eq70}) we easily obtain
\begin{equation}\label{eq80}
\hat{S}(\chi) | \bx \ra  =  e^{\im f(\chi, \phi)} | \bx \ra,
\end{equation}
which shows, together with Eq. (\ref{eq60}), that the SPP operator
$\hat{S}(\chi)$ is symmetric.
From Eq. (\ref{eq80}) it is straightforward to write the
corresponding transformation law for the spatial creation
operators $\hat{a}^\dagger(\bx)$:
\begin{equation}\label{eq85}
\hat{S}(\chi) \hat{a}^\dagger(\bx) \hat{S}^\dagger (\chi)= e^{\im
f(\chi, \phi)} \hat{a}^\dagger(\bx).
\end{equation}

To conclude this paragraph, we define the complementary spiral
phase plate, as a SPP that produces  a negative
azimuthal-dependent phase shift $\exp(-\im f(\chi, \phi))$ on a
crossing beam:
\begin{equation}\label{eq90}
\psi(\bx) \rightarrow e^{-\im f(\chi, \phi)} \psi(\bx).
\end{equation}
Then from  the Hermitian-conjugate of Eq. (\ref{eq60}) it readily
follows that the CSPP is represented by  $\hat{S}^\dagger(\chi)$.
%
%
%
%
%
\subsection{The Mach-Zehnder interferometer}
Figure 2) shows a detailed scheme of the Mach-Zehnder
interferometer $\mathrm{MZ}_a$. Photon $a$ enters the Mach-Zehnder
through the channel ``$1$'' and interacts with the first $50/50$
beam splitter. Channel ``$2$'' is fed with vacuum.
A
general two-mode single-photon (TMSP) state at the input of
$\mathrm{MZ}_a$ can be therefore written
%
%
%
\begin{equation}\label{eq100}
\begin{array}{rcl}
\displaystyle{|\psi_a \ra} & = & \displaystyle{\int \mathrm{d}^2
\bx \, \xi(\bx)\hat{a}_1^\dagger(\bx) | 0 \ra},
\end{array}
\end{equation}
where the subscript  $1$ denotes the channel $1$.
 The beam splitter transforms the input TMSP state $| \psi_a
 \ra$ in a superposition of TMSP states \cite{Campos89a}, $|\psi_a \ra \rightarrow |\psi_a
 \ra_1$:
\begin{equation}\label{eq110}
\begin{array}{rcl}
\displaystyle{|\psi_a \ra_1} & = &  \displaystyle{\hat{U}^\dagger
|\psi_a
\ra}\\\\
 & = & \displaystyle{\int
\mathrm{d}^2 \bx \, \xi(\bx) \hat{U}^\dagger\hat{a}_1^\dagger(\bx)
\hat{U}| 0 \ra},
\end{array}
\end{equation}
where \cite{Walborn03a,Dang05a}
\begin{equation}\label{eq114}
\begin{array}{rcl}
\displaystyle{\hat{U}^\dagger\hat{a}_1^\dagger(\bx) \hat{U}} & = &
\displaystyle{t \hat{a}_1^\dagger(x,y) - r^* \hat{a}_2^\dagger(x,-y)},\\\\
\displaystyle{\hat{U}^\dagger\hat{a}_2^\dagger(\bx) \hat{U}} & = &
\displaystyle{r \hat{a}_1^\dagger(x,-y) +t^*
\hat{a}_2^\dagger(x,y)},
\end{array}
\end{equation}
 and we assume
$t = 1/\sqrt{2}, \, r = \im / \sqrt{2}$ for the transmission and
reflection coefficients of the $50/50$ beam splitter,
respectively. We can then write
\begin{equation}\label{eq116}
\begin{array}{rcl}
\displaystyle{|\psi_a \ra_1} & = &  \displaystyle{\int
\mathrm{d}^2 \bx \, \frac{\xi(\bx)}{\sqrt{2}}\left[
\hat{a}_1^\dagger(x,y)| 0 \ra + \im \hat{a}_2^\dagger(x,-y)| 0 \ra
\right]}.
\end{array}
\end{equation}
The action of the mirror $\mathrm{M}_2$ can be described as
$\hat{a}_2^\dagger(x,-y) \rightarrow r \hat{a}_2^\dagger(x,y)$,
where $r = \im$, which causes $|\psi_a \ra_1 \rightarrow |\psi_a
\ra_2$, where
\begin{equation}\label{eq117}
\begin{array}{rcl}
\displaystyle{|\psi_a \ra_2} & = &  \displaystyle{\int
\mathrm{d}^2 \bx \, \frac{\xi(\bx)}{\sqrt{2}}\left[
\hat{a}_1^\dagger(x,y)| 0 \ra - \hat{a}_2^\dagger(x,y)| 0 \ra
\right]}.
\end{array}
\end{equation}
After the first beam splitter and the mirror $\mathrm{M}_2$, there
are two SPPs, one per channel, which perform a unitary operation
on the TMSP state: $|\psi_a \ra_2 \rightarrow |\psi_a \ra_3$. Let
$\hat{S}(\alpha_i)$ the operator representing the SPP (CSPP) in
the channel $i$ ($i=1,2$). Since under $\hat{S}(\alpha_i)$, the
operators $\hat{a}_i^\dagger(\bx)$ transform as
\begin{equation}\label{eq118}
\hat{a}_i^\dagger(\bx) \rightarrow \hat{S}(\alpha_i)
\hat{a}_i^\dagger(\bx)\hat{S}^\dagger(\alpha_i) = e^{\im
f{(\alpha_i,\phi)}} \hat{a}_i^\dagger(\bx),
\end{equation}
then the TMSP state after the SPPs can be written as
\begin{equation}\label{eq120}
\displaystyle{|\psi_a \ra_3 = \int \mathrm{d}^2 \bx \,
\frac{\xi(\bx)}{\sqrt{2}}\left[ e^{\im f(\alpha_1,\phi)}
\hat{a}_1^\dagger(\bx)| 0 \ra - e^{\im f(\alpha_2,\phi)}
\hat{a}_2^\dagger(\bx)| 0 \ra \right]},
\end{equation}
where Eq. (\ref{eq80}) has been used. The action of the mirror
$\mathrm{M}_1$ can be described as $\hat{a}_1^\dagger(x,y)
\rightarrow -r^* \hat{a}_1^\dagger(x,-y)$, where $r = \im$, which
causes $|\psi_a \ra_3 \rightarrow |\psi_a \ra_4$, where
\begin{equation}\label{eq125}
\begin{array}{rclcl}
\displaystyle{|\psi_a \ra_4} & = & \displaystyle{ \int
\mathrm{d}^2 \bx \, \frac{\xi(\bx)}{\sqrt{2}}} &
\displaystyle{\Bigl[ }& \displaystyle{\im e^{\im f(\alpha_1,\phi)}
\hat{a}_1^\dagger(x,-y)| 0 \ra}\\ && && \displaystyle{ - e^{\im
f(\alpha_2,\phi)} \hat{a}_2^\dagger(x,y)| 0 \ra \Bigr]}.
\end{array}
\end{equation}
 The second beam splitter transforms the creation operators according to \cite{Walborn03a}
\begin{equation}\label{eq130}
\begin{array}{rcl}
\displaystyle{\hat{a}_1^\dagger(x,-y)} & \rightarrow &
\displaystyle{r_a \hat{a}_2^\dagger(x,y) +t_a^*
\hat{a}_1^\dagger(x,-y)},\\\\
\displaystyle{\hat{a}_2^\dagger(x,y)} & \rightarrow &
\displaystyle{t_a \hat{a}_2^\dagger(x,y) - r_a^*
\hat{a}_1^\dagger(x,-y)},
\end{array}
\end{equation}
where $t_a = \cos \theta_a, \; r_a = \im \sin \theta_a$.
 The TMSP state $|\psi_a \ra_5 $ at the output of the
Mach-Zehnder $\mathrm{MZ}_a$ can therefore be written as
\begin{equation}\label{eq140}
\begin{array}{rcl}
\displaystyle{|\psi_a \ra_5} & = & \displaystyle{ \int
\mathrm{d}^2 \bx \, \frac{\xi(\bx)}{\sqrt{2}} \Bigl\{ \left[\im
r_a e^{\im
f(\alpha_1,\phi)}  - t_a e^{\im f(\alpha_2,\phi)}\right]\hat{a}_2^\dagger(x,y)}\\\\
&& \displaystyle{ + \left[\im t_a^* e^{\im f(\alpha_1,\phi)}
+r_a^* e^{\im f(\alpha_2,\phi)}\right]\hat{a}_1^\dagger(x,-y)}
\Bigr\} | 0 \ra.
\end{array}
\end{equation}
The role of the last mirror $\mathrm{M}_3$ is to maintain the
initial spatial wave-function  $\xi (\bx)$ invariant; its action
can be described as $\hat{a}_1^\dagger(x,-y) \rightarrow r
\hat{a}_1^\dagger(x,y)$, $(r = \im)$, which causes $|\psi_a \ra_5
\rightarrow |\psi_a \ra_6$, where, apart an overall phase factor,
\begin{equation}\label{eq145}
\begin{array}{rcl}
\displaystyle{|\psi_a \ra_6} & = & \displaystyle{ \int
\mathrm{d}^2 \bx \, \frac{\xi(\bx)}{\sqrt{2}} \Bigl\{  \left[
t_a^* e^{\im f(\alpha_1,\phi)}  - \im r_a^* e^{\im
f(\alpha_2,\phi)}\right]\hat{a}_1^\dagger(x,y)}\\\\
&& \displaystyle{+ \left[-\im r_a e^{\im
f(\alpha_1,\phi)}  + t_a e^{\im f(\alpha_2,\phi)}\right]\hat{a}_2^\dagger(x,y) \Bigr\} | 0 \ra }\\\\
& = & \displaystyle{ \int \mathrm{d}^2 \bx \,
\frac{\xi(\bx)}{\sqrt{2}} \Bigl\{ }\\\\
&& \displaystyle{  \; \; \;\left[ \cos \theta_a e^{\im
f(\alpha_1,\phi)} -\sin \theta_a e^{\im
f(\alpha_2,\phi)}\right]\hat{a}_1^\dagger(x,y)}\\\\
&& \displaystyle{+ \left[\sin \theta_a e^{\im
f(\alpha_1,\phi)}  + \cos \theta_a e^{\im f(\alpha_2,\phi)}\right]\hat{a}_2^\dagger(x,y) \Bigr\} | 0 \ra }\\\\
& \equiv & \displaystyle{\int \mathrm{d}^2 \bx \, \xi(\bx) \left[
A_1(\phi) \hat{a}_1^\dagger(\bx) + A_2(\phi)
\hat{a}_2^\dagger(\bx) \right] | 0 \ra}.
\end{array}
\end{equation}
%
%
%
Equation (\ref{eq145}) shows that for a given $\bx$ the TMSP state
$|\psi_a \ra_6$ spans, as $\theta_a$ varies, a two-dimensional
space determined by the orthogonal basis
$\{\hat{a}_1^\dagger(\bx)| 0\ra,\hat{a}_2^\dagger(\bx)| 0\ra \}$.
In fact, if we define
\begin{equation}\label{eq152}
\vec{A}(\phi) = \begin{pmatrix}
  A_1 \\
  A_2
\end{pmatrix}, \quad \vec{E}(\phi) =
\frac{1}{\sqrt{2}}\begin{pmatrix}
  e^{\im f_1} \\
  e^{\im f_2}
\end{pmatrix},
\end{equation}
where $f_i \equiv f(\alpha_i,\phi)$, then from Eq. (\ref{eq145})
it readily follows
\begin{equation}\label{eq154}
\vec{A}(\phi) = \textsf{R}(\theta_a) \vec{E}(\phi),
\end{equation}
where
\begin{equation}\label{eq156}
\textsf{R}(\theta_a) = \begin{pmatrix}
  \cos \theta_a & -\sin \theta_a \\
  \sin \theta_a & \cos \theta_a
\end{pmatrix},
\end{equation}
is the well known $2 \times 2$ rotation matrix. Therefore, as
$\theta_a$ varies from $0$ to $2\pi$, the state $|\psi_a \ra_6$
makes a complete rotation in the plane $\{\hat{a}_1^\dagger(\bx)|
0\ra,\hat{a}_2^\dagger(\bx)| 0\ra \}$. It is easy to see that if
we take in account the azimuthal-independent phases
$\varphi_{xi}$, the orthogonal matrix $\textsf{R}(\theta_a)$ must
be replaced by the unitary matrix $\textsf{U}(\theta_a)$ defined
as
\begin{equation}\label{eq157}
\textsf{U}(\theta_a) = \begin{pmatrix}
 e^{\im \varphi_{a1}}  \cos \theta_a & - e^{\im \varphi_{a2}} \sin \theta_a\\
 e^{\im \varphi_{a1}}   \sin \theta_a & e^{\im \varphi_{a2}} \cos \theta_a
\end{pmatrix},
\end{equation}

 Now we can repeat for the photon $b$ the very
same calculation beginning with the state
\begin{equation}\label{eq158}
\begin{array}{rcl}
\displaystyle{|\psi_b \ra} & = & \displaystyle{\int \mathrm{d}^2
\bx \, \xi(\bx)\hat{b}_1^\dagger(\bx) | 0 \ra},
\end{array}
\end{equation}
 at the input of the Mach-Zehnder $\mathrm{MZ}_b$ and ending with the state
$|\psi_b\ra_6$ at the output of $\mathrm{MZ}_b$:
\begin{equation}\label{eq160}
|\psi_b \ra_6 =  \int \mathrm{d}^2 \bx \, \xi(\bx) \left[
B_1(\phi) \hat{b}_1^\dagger(\bx) + B_2(\phi)
\hat{b}_2^\dagger(\bx) \right] | 0 \ra,
\end{equation}
where
\begin{equation}\label{eq170}
\begin{array}{rcl}
\displaystyle{B_1(\phi)} & = & \displaystyle{[ \cos \theta_a
e^{-\im f(\beta_1, \phi)} -\sin \theta_a e^{-\im
f(\beta_2, \phi)}]/\sqrt{2}},\\\\
\displaystyle{B_2(\phi)} & = & \displaystyle{ [\sin \theta_a
e^{-\im f(\beta_1, \phi)}  + \cos \theta_a e^{-\im f(\beta_2,
\phi)}]/\sqrt{2}}.
\end{array}
\end{equation}
 Note that the minus sign in the exponentials is due to the fact that
CSPPs (instead of SPP) are used in the Mach-Zehnder
$\mathrm{MZ}_b$.

Since the $A_i(\phi)$ and $B_i(\phi)$ does not depend on the
radial coordinate $r = |\bx|$, in the following we shall indicate
them as $A_i(\phi)$ and $B_i(\phi)$, respectively.
\subsection{The twin-photon state}
The OAM-entangled  state of a photon pair emitted by a crystal
pumped by a $\mathrm{LG}_0^0$ laser beam, can be written
\cite{Visser04a}
\begin{equation}\label{eq190}
| \Psi \ra \propto \int \mathrm{d}^2 \bx \, \Lambda_P(r)
\hat{a}^\dagger(\bx) \hat{b}^\dagger(\bx)| 0 \ra,
\end{equation}
where $\Lambda_P(r) \equiv \mathrm{LG}_0^0(r,w_P)$  describes the
transverse profile of the pump beam,  $r = |\bx|$, and $w_P$ is
the beam waist.  The state $| \Psi \ra$ is clearly
non-normalizable, therefore we use the symbol ``$\propto$''
instead of ``$=$''. As we shall see, this fact does not represent
a problem since all the measurable probabilities will be properly
normalized. Passing from the single-mode to the two-mode
description introduced in the previous paragraph, we rewrite the
two-mode two-photon (TMTP) state  $| \Psi \ra$ as
\begin{equation}\label{eq200}
\begin{array}{rcl}
| \Psi \ra & \propto & \displaystyle{ \int \mathrm{d}^2 \bx \,
\Lambda_P(r)  \hat{a}_1^\dagger(\bx) \hat{b}_1^\dagger(\bx)| 0
\ra}
%
\end{array}
\end{equation}
to indicate that both photons enter the channel ``1'' of their
respective interferometers. When the photon pair crosses both the
Mach-Zehnder, $| \Psi \ra$ undergoes the transformation $| \Psi
\ra \rightarrow | \Psi' \ra$:
\begin{equation}\label{eq210}
\begin{array}{rcl}
| \Psi' \ra & \propto & \displaystyle{ \int \mathrm{d}^2 \bx \,
\Lambda_P(r) \Bigl\{ \left[ A_1(\phi) \hat{a}_1^\dagger(\bx) +
A_2(\phi) \hat{a}_2^\dagger(\bx)
\right]| 0 \ra}\\\\
&  & \displaystyle{ \otimes \left[ B_1(\phi)
\hat{b}_1^\dagger(\bx) + B_2(\phi) \hat{b}_2^\dagger(\bx)
\right] | 0 \ra \Bigr\}}\\\\
&  & \displaystyle{ = \sum_{i,j}^{1,2}\int \mathrm{d}^2 \bx \,
\Lambda_P(r) \Psi_{ij}(\phi) | \bx\ra_{ai}  | \bx \ra_{bj} },
\end{array}
\end{equation}
where $| \bx\ra_{ai}  | \bx \ra_{bj}  \equiv
\hat{a}_i^\dagger(\bx)  \hat{b}_j^\dagger(\bx)| 0 \ra $ denotes a
position state with the photon $a$ in the channel $i$ and the
photon $b$ in the channel $j$, and
\begin{equation}\label{eq222}
\Psi_{ij}(\phi) = \varsigma_{ij} A_i(\phi) B_j(\phi), \qquad
(i,j=1,2),
\end{equation}
and $\varsigma_{ij} = (3 - i -j) +\im( 3i + 3j - 2ij - 4 )$.
\subsection{The single-mode fibers}
Figure 2) shows that the output channel $i, \; (i =1,2)$ of each
Mach-Zehnder $\mathrm{MZ}_x, \; (x =a,b)$ is coupled with the
single-mode fiber $\mathrm{F}_{xi}$ which sustains the
Laguerre-Gaussian mode $\mathrm{LG}_0^0$ with waist $w_0$. For a
proper quantum mechanical description of the fiber we need to
introduce the Laguerre-Gaussian single-photon states defined as
\begin{equation}\label{eq230}
|l, p\ra = \int \mathrm{d}^2 \bx \, \mathrm{LG}_p^l(\bx) | \bx\ra.
\end{equation}
From the orthogonality property of the LG functions, it readily
follows
\begin{equation}\label{eq240}
\begin{array}{rcl}
\la l, p|l', p' \ra &= & \displaystyle{\int \mathrm{d}^2 \bx \,
\left[\mathrm{LG}_p^l(\bx)\right]^*  \mathrm{LG}_{p'}^{l'}(\bx) }\\\\
& =& \displaystyle{\delta_{ll'} \delta_{pp'}}.
\end{array}
\end{equation}
When a photon in the arbitrary  state $|\xi \ra$  is coupled to a
single-mode fiber, the fiber transforms the input state of the
photon in the Laguerre-Gaussian state $ | l=0, p=0\ra
\equiv|0,0\ra$ with probability $|\la 0,0 | \xi \ra|^2$. Since in
our scheme the output port of each single-mode fiber is coupled to
a detector, the probability $P_{ij}(\theta_a, \theta_b)$ that the
detector $\mathrm{D}_{ai}$ fires in coincidence with the detector
 $\mathrm{D}_{bj}$ is given by
\begin{equation}\label{eq250}
\begin{array}{rcl}
\displaystyle{ P_{ij}(\theta_a, \theta_b)} & \propto &
\displaystyle{ \left| \int \mathrm{d}^2 \bx \, \Lambda_P(r)
\Psi_{ij}(\phi) \la 0,0|\bx\ra_{ai} \la 0,0| \bx \ra_{bj}
 \right|^2 }\\
 &  &
\displaystyle{= \left| \int \mathrm{d}^2 \bx \, \Lambda_P(r)
 \Lambda_0^2(r) \Psi_{ij}(\phi)
 \right|^2 },
 \end{array}
\end{equation}
since $\la 0,0|\bx\ra_{ai} =\la 0,0| \bx \ra_{bj}=\la 0,0| \bx
\ra$, and
\begin{equation}\label{eq260}
\la 0,0|\bx\ra = \mathrm{LG}_0^0(r, w_0) = \sqrt{\frac{2}{\pi
w_0^2}} e^{-r^2/w_0^2} \equiv \Lambda_0(r).
\end{equation}
As the $ \Psi_{ij}(\phi)$s do not depend on $r$, we can factorize
$P_{ij}(\theta_a, \theta_b)$ by passing to polar coordinates
$(x,y) \rightarrow (r, \phi)$:
\begin{equation}\label{eq270}
\begin{array}{l}
\displaystyle{\int \mathrm{d}^2 \bx \, \Lambda_P(r)
 \Lambda_0^2(r) \Psi_{ij}(\phi)}\\\\  \displaystyle{ \; \; \; \; \;\; \; \; \; \; =
 \int_0^\infty \mathrm{d} r \, r\Lambda_P(r)
 \Lambda_0^2(r)  \int_0^{2 \pi} \mathrm{d}\phi \Psi_{ij}(\phi)},
  \end{array}
\end{equation}
and cast the radial part aside in order to get
\begin{equation}\label{eq280}
P_{ij}(\theta_a, \theta_b) \propto \left| \int_0^{2 \pi}
\mathrm{d}\phi \Psi_{ij}(\phi) \right|^2.
\end{equation}

Finally, from Eqs. (\ref{eq30},\ref{eq140}-\ref{eq170}) we note
that, in practice, the only elementary azimuthal integral one
needs to calculate is
\begin{equation}\label{eq300}
\begin{array}{rcl}
\displaystyle{ I(\mu, \nu, \mathcal{L}) } & = & \displaystyle{
\int_0^{2 \pi} \mathrm{d} \phi e^{\im \left[ f(\phi,\mu) - f(\phi,
\nu) \right]}}  \\\\
& = & \displaystyle{ e^{\im(\mu - \nu)\mathcal{L}} \Bigl\{ 2 \pi -  \Bigl[ (1 - e^{\im 2 \pi \mathcal{L}})\Theta(\mu -\nu)}\\\\
&&\displaystyle{ - (1 - e^{-\im 2 \pi \mathcal{L}})\Theta(\nu
-\mu) \Bigr](\mu - \nu)
 \Bigr\}  },
  \end{array}
\end{equation}
which reduces to the simpler form
\begin{equation}\label{eq310}
\displaystyle{ I(\mu, \nu, l + 1/2) } = 2 \pi e^{- \im(\mu -
\nu)(l + 1/2)} \left( 1 - \frac{|\mu - \nu|}{\pi} \right),
\end{equation}
for $\mathcal{L} = l + 1/2$, where $l \in \{ 0, 1, 2, \ldots \} $.
\section{The Clauser-Horne inequality}
In the previous section we calculated directly the coincidence
probabilities $P_{ij}(\theta_a,\theta_b)$ from the TMTP state $|
\Psi'\ra$ at the output of both interferometers. However, to
proceed further and test the non-locality of the  state $|
\Psi'\ra$, we have to specify our scenario more precisely. Let us
formalize our experiment as follows. There are two parties, say
Alice and Bob, who share the two-photon entangled state $| \Psi
\ra$ given in Eq. (\ref{eq190}). Each one of the two entangled
photons belong to an $\infty$-dimensional Hilbert space, namely
the OAM Hilbert space. Alice and Bob have two distinct measuring
apparatuses: $\mathrm{M}_a$ and $\mathrm{M}_b$ respectively. Each
apparatus $\mathrm{M}_x, \; (x=a,b)$ consists of a two-channel
Mach-Zehnder interferometer $\mathrm{MZ}_x$, with a parameter
$\theta_x$ at the experimenter's disposal, followed by two (one
per channel) single-mode fibers $\mathrm{F}_{xi}, \; (i=1,2)$. The
output ports $i =1,2$ of each $\mathrm{M}_x$ are monitored by two
detectors $\mathrm{D}_{x1}$ and $\mathrm{D}_{x2}$ respectively. We
stress that in \emph{this} scenario the SPPs rotation angles
$\alpha$ and $\beta$ are \emph{not} experimental ``knobs''
 that can be changed during an experiment.
Different pairs $\{ \alpha,\beta \}$ define  \emph{different}
experiments which use the \emph{same} initial two-photon entangled
state $| \Psi\ra$. In analogy with the polarization case, Alice
can choose between $2$ different measurements, say $A$ and $A'$,
corresponding to two different choices for the
varying-beam-splitter ``angles'' $\theta_a$ and $\theta_a'$,
respectively. Similarly, Bob can choose between  $B$ and $B'$,
corresponding to $\theta_b$ and $\theta_b'$, respectively. Each
time Alice and Bob perform a measurement, $\mathrm{M}_x$ $(x=a,b)$
gives the string $\{x_1, x_2\}$, where $ x_i = 1$ when the
detector $\mathrm{D}_{xi}$ fires and $ x_i = 0$ when it does not.
At this point, our scenario is completely defined: We have $2$
parties (Alice and Bob), $2$ measurements ($\theta_x$ and
$\theta_x'$) per party, and $2$ possible outcomes ($\{1, 0\}$ and
$\{0, 1\}$) per measurement per party.

Now we can calculate  the quantum mechanical predictions for the
experimental outcomes. These calculations were already done in the
previous paragraph, but here we want to repeat them in a slightly
different way in order to display the dichotomic nature of the
problem.
To begin with, we fix for the rest of this paper, $\alpha_1 \equiv
\alpha, \; \alpha_2 = \alpha + \pi $ and $\beta_1 \equiv \beta, \;
\beta_2 = \beta + \pi $. Moreover, we fix $\mathcal{L} = l + 1/2$,
where $l \in \{ 0, 1, 2, \ldots \}$. The TMTP state $| \Psi'\ra$
 (\ref{eq210}) describes the photon pair at the output the two
interferometers, just before the fibers. As shown previously, each
fiber projects any input single-photon state in the
Laguerre-Gaussian state $|0,0\ra$. Therefore, from Eq.
(\ref{eq210}) it readily follows that the two-photon state $|
\Psi''\ra$ after the fibers can be written as
\begin{equation}\label{eq320}
\begin{array}{rcl}
| \Psi'' \ra & \propto & \displaystyle{ \sum_{i,j}^{1,2}
|0,0\ra_{ai}|0,0\ra_{bj} }\\\\
&& \displaystyle{ \times \int \mathrm{d}^2 \bx \, \Lambda_P(r)
\Psi_{ij}(\phi) \la 0,0|\bx \ra_{ai} \la 0,0|\bx \ra_{bj} }.
\end{array}
\end{equation}
As was shown in Eq. (\ref{eq270}), it is possible to write
\begin{equation}\label{eq330}
\begin{array}{l}
\displaystyle{ \int \mathrm{d}^2 \bx \, \Lambda_P(r) \Psi_{ij}(\phi) \la 0,0|\bx \ra_{ai} \la 0,0|\bx \ra_{bj} }\\\\
 \displaystyle{\; \; \; =  \int_0^\infty \mathrm{d} r \, r \Lambda_P(r)\Lambda_0^2(r)
\times \int_0^{2 \pi} \mathrm{d} \phi \, \Psi_{ij}
(\phi)}\\\\
\displaystyle{ \; \; \; \equiv R \times C_{ij}(\theta_a,
\theta_b)},
\end{array}
\end{equation}
where the radial integral $R$ does not depend nor on $\alpha$ and
$\beta$, nor on $\theta_a$ and $\theta_b$. We can write then
\begin{equation}\label{eq340}
| \Psi'' \ra \propto \sum_{i,j}^{1,2} C_{ij}|i,j\ra,
\end{equation}
where $R$ has been absorbed into the proportionality factor and
$|i,j\ra$ is a shorthand for $|0,0\ra_{ai}|0,0\ra_{bj}$. At this
point, it is straightforward to write the normalized TMTP state
$|\Psi_{00}\ra$ after the fibers as
\begin{equation}\label{eq350}
| \Psi_{00} \ra = \sum_{i,j}^{1,2} \lambda_{ij}|i,j\ra,
\end{equation}
were we have defined the two-photon amplitudes
\begin{equation}\label{eq360}
 \lambda_{ij}(\theta_a, \theta_b)= \frac{\varsigma_{ij} \int_0^{2 \pi}\mathrm{d} \phi \, A_{i}(\phi)  B_{j}
(\phi) }{\sqrt{\sum_{i,j}^{1,2} \left| \int_0^{2 \pi} \mathrm{d}
\phi \, A_{i}(\phi) B_{j} (\phi)  \right|^{2}}}.
\end{equation}
The state $|\Psi_{00}\ra$ is clearly entangled since the
coefficients $ \lambda_{ij}$ are, in general, not factorable.
Moreover, it belongs to a $4$-dimensional Hilbert space, as a
two-photon polarization-entangled state, since the continuous
variables $(r, \phi)$ have been integrated out. In fact, all our
operations can be summarized in this way: We began with an
OAM-entangled two-photon state belonging to an
infinite-dimensional Hilbert space $\mathcal{H}_{ab}^{\infty
\times \infty}$. Then we performed on this state some unitary
operations which permitted us to span a certain sub-space of
$\mathcal{H}_{ab}^{\infty \times \infty}$. Finally, we projected
the transformed state onto a $4$-dimensional Hilbert space
$\mathcal{H}_{ab}^{2 \times 2}$, the two dimensions (per photon)
being provided by the two spatial modes (``arms'') of the
Mach-Zehnder interferometer. In this way the
entanglement-preserving mapping $\mathcal{H}_{ab}^{\infty \times
\infty} \rightarrow \mathcal{H}_{ab}^{2 \times 2}$ was
accomplished. We stress that the azimuthal integration in Eq.
(\ref{eq360}) clearly shows that the final state $| \Psi_{00}\ra$
is entangled because the initial state $| \Psi \ra$ from the
crystal was entangled, and not because the beam splitters in the
MZs \emph{created} the entanglement \cite{Kim02a}.

Now that we have reduced our problem to a $4$-dimensional one,
there are several inequalities at our disposal to check the
non-locality of the state $|\Psi_{00}\ra$. The best known are the
Bell inequality \cite{Bell1}, the Clauser-Horne-Shimony-Holt
(CHSH) inequality \cite{Clauser69}, and the  Clauser-Horne (CH)
inequality \cite{Clauser74a}. Since we are proposing an
experiment, we choose here to check the CH inequality  which,
differently from the CHSH inequality, does not require the fair
sampling hypothesis \cite{GaruccioeRapisarda} to allow the use of
unnormalized experimental data.
 In practice, an experimenter choose a measurement, say
 $(A,B)$, and repeats it $N$ times ($N$ realizations) obtaining
 two strings $\{x_{1k}, x_{2k} \}, \; (x=a,b; \; x_{ik}=0,1)$ for each realization $k, \;( k = 1, \ldots,
 N)$. Then, for $N \gg 1$, the coincidence probabilities $P_{ij}(\theta_a,
 \theta_b)$ are well approximated by the coincidence frequencies $F_{ij}(\theta_a,
 \theta_b)$
\begin{equation}\label{eq370}
F_{ij}(\theta_a, \theta_b) = \frac{1}{N} \sum_{k=1}^N
\Theta(a_{ik} b_{jk} -1/2 ),
\end{equation}
where $\Theta$ is the Heaviside step function. These frequencies
 are clearly not ``absolute'' since in a real experiment there are
always missing outcomes due, for example,  to detector
inefficiencies and to losses. In other words, we can say that the
experimenter has not access to the normalized state $| \Psi_{00}
\ra$, but only to the unnormalized one $| \Psi'' \ra$
(\ref{eq340}).
Therefore, in order write the CH inequality in a useful form for
an experimenter, we  calculate the following unnormalized
coincidences probabilities
\begin{subequations}
\begin{eqnarray}
\displaystyle{ P_{ab}(\theta_a, \theta_b) } & = & \displaystyle{ p_{11} ,}\label{eq380a}\\
\displaystyle{ P_{ab}(\theta_a, \infty)} & = & \displaystyle{
p_{11}  +
  p_{12} ,}\label{eq380b}\\
\displaystyle{ P_{ab}(\infty, \theta_b)} & = & \displaystyle{
p_{11}  +
 p_{21} ,}\label{eq380c}\\
\displaystyle{ P_{ab}(\infty, \infty)} & = & \displaystyle{
 p_{11}  +
 p_{12} +
 p_{21}  +
 p_{22} },\label{eq380d}
\end{eqnarray}
\end{subequations}
where $p_{ij}  = \left| C_{ij}(\theta_a, \theta_b)\right|^2$, and
define the Bell-Clauser-Horne parameter  $S$ as
\begin{widetext}
\begin{equation}\label{eq390}
S = \frac{ P_{ab}(\theta_a, \theta_b)- P_{ab}(\theta_a,
\theta_b')+P_{ab}(\theta_a', \theta_b)+P_{ab}(\theta_a',
\theta_b')-P_{ab}(\theta_a', \infty) - P_{ab}(\infty,
\theta_b)}{P_{ab}(\infty, \infty) }.
\end{equation}
\end{widetext}
Then, the CH inequality requires
\begin{equation}\label{eq400}
S \leq 0,
\end{equation}
for any objective local theory.

%
%
From Eq. (\ref{eq380a}-\ref{eq380d}) it is simple to calculate the
four coincidence probabilities: They are explicitly given in
Appendix A. They seems complicated but after a careful inspection
it is easy to see that if we choose a common orientation $\alpha =
\beta$ for the SPPs and the CSPPs for the two photons, they reduce
to the simpler form
\begin{subequations}
\begin{eqnarray}
\displaystyle{ \frac{P_{ab}(\theta_a, \theta_b)}{P_{ab}(\infty, \infty)} }
& = & \displaystyle{ \frac{1}{2} \cos^2(\theta_a - \theta_b) ,}\label{eq400a}\\
\displaystyle{ \frac{P_{ab}(\theta_a, \infty)}{P_{ab}(\infty,
\infty)}} & = & \displaystyle{\frac{1}{2} ,}\label{eq400b}\\
\displaystyle{\frac{ P_{ab}(\infty, \theta_b)}{P_{ab}(\infty,
\infty)}} & = & \displaystyle{\frac{1}{2} .}\label{eq400c}
\end{eqnarray}
\end{subequations}
With the particular choice of varying-beam-splitter angles
\[
\begin{array}{lcl}
  \theta_a & = & 0, \\
  \theta_a' & = & \pi/4, \\
  \theta_b & = & \pi/8, \\
  \theta_b' & = & 3 \pi/8,
\end{array}
\]
we achieve the maximum violation $S = (\sqrt{2}-1)/2$ of the CH
inequality. This result is valid for \emph{all} pairs of
``external'' parameters $(\alpha, \beta = \alpha)$.

This is the main result of this paper. Differently from  the
polarization case, here we have the additional parameter $\alpha$
which can be varied from $0$ to $2 \pi$ in order to span part of
the infinite-dimensional OAM-entangled two-photon Hilbert space.
Different values of $\alpha$ define different experiments and
\emph{all} these experiments give the maximum violation of the CH
inequality.

We stress that the condition $\alpha = \beta$ is sufficient but
not necessary to obtain high violation of CH inequality in our
scheme. In fact, by numerical search, we found many pairs $\alpha
\neq \beta $ which produces violations bigger than, e.g.,
$0.204$.
%
 %
 %
 %
 %
%
\section{Discussion and conclusions}
What is the meaning of the SPP orientation angles pair $(\alpha,
\beta)$? In order to answer this question, let  us summarize our
previous results as follows. Consider a detection event
represented by the string $D_{ij}(\theta_a,\theta_b) \equiv \{ \{
a_1, a_2\},\{b_1,b_2 \} \}$. It is not difficult to show that the
probability $P_{ij}(\theta_a,\theta_b)$ of such an event can be
written as
\begin{equation}\label{eq320a}
P_{ij}(\theta_a,\theta_b) = \bigl| \la0,0| \la0,0|
\hat{U}_i(\alpha,\theta_a) \otimes
\hat{U}_j^\dagger(\beta,\theta_b) | \Psi^\mathrm{in} \ra \bigr|^2,
\end{equation}
where
\begin{equation}\label{eq330a}
\hat{U}_i(\chi,\theta_x) = \sum_{j=1}^2
R_{ij}(\theta_x)\hat{S}(\chi_j)/ \sqrt{2}, \, \quad
\left\{\begin{array}{c}
  i=1,2, \\
  x = a,b,
\end{array}\right.
\end{equation}
$[\chi_j = \chi + (j-1)\pi, \, (\chi = \alpha, \beta)]$ is the
operator representing the propagation of a photon through the
channel ``$i$'' of $\mathrm{MZ}_x$, and $\hat{S}(\chi_j)$ is the
quantum-mechanical operator representing a SPP oriented at angle
$\chi_j$ \cite{Aiello05d}. From Eqs. (\ref{eq320}-\ref{eq330}) it
follows that when the event $D_{ij}(\theta_a,\theta_b)$ occurs,
the input state $| \Psi^\mathrm{in} \ra$ is projected onto the
state $| u_i(\alpha,\theta_a) \ra| \bar{u}_j(\beta,\theta_b) \ra$,
where
\begin{equation}\label{eq340a}
\begin{array}{rcl}
\displaystyle{ | u_i(\alpha,\theta_a) \ra} & =&
\displaystyle{\hat{U}_i^\dagger(\alpha,\theta_a)|0,0\ra } \\
& \equiv & \displaystyle{\sum_{j=1}^2
R_{ij}(\theta_a)|S(\alpha_j)\ra/ \sqrt{2} },
  \end{array}
\end{equation}
and $|S(\alpha_j)\ra \equiv \hat{S}^\dagger(\alpha_j)|0,0\ra$. In
a similar manner we define   $| \bar{u}_j(\beta,\theta_b)\ra =
\hat{U}_j(\beta,\theta_b)|0,0\ra $ and $|\bar{S}(\beta_j)\ra
\equiv \hat{S}(\beta_j)|0,0\ra$. From
 $\la S(\alpha_i)|S(\alpha_j)\ra = \delta_{ij}
=\la \bar{S}(\beta_i)|\bar{S}(\beta_j)\ra $ \cite{Sumant04a}, it
follows that
 $\{ |S(\alpha)\ra, |S(\alpha+\pi) \ra \}$ and
$\{ |\bar{S}(\beta)\ra, |\bar{S}(\beta +\pi) \ra \}$ form an
orthogonal two-dimensional basis for the photons $a$ and $b$,
respectively. Therefore,
 Eq. (\ref{eq340}) tells us that the  state $|
u_i(\alpha,\theta_a) \ra| \bar{u}_j(\beta,\theta_b) \ra$ onto
which Alice and Bob project their state $| \Psi^\mathrm{in} \ra$,
is confined to the four-dimensional two-photon subspace spanned by
the basis $\{ |S(\alpha_i)\ra \otimes |\bar{S}(\beta_j)\ra \}, \,
(i,j=1,2)$. Moreover, we can see that, e.g., the basis $\{
|S(\alpha)\ra, |S(\alpha+\pi) \ra \}$ defines a dichotomic
subspace as the basis $\{ |H \ra, |V \ra \}$ does in the
polarization space.
It is clear then that when we choose a pair $(\alpha, \beta)$ of
SPPs orientations, we uniquely fix a four-dimensional two-photon
subspace.
Since the CH inequalities are applicable to the counts at any pair
of detectors which measure dichotomic variables (irrespective of
the dimensionality of the bipartite quantum state $|
\Psi^\mathrm{in} \ra$ under examination \cite{Peres78a,Zuko97a}),
we can choose other pairs $(\alpha', \beta')$ of SPPs orientations
(which define \emph{other} four-dimensional  two-photon
subspaces), repeat the  measurements, and find again the maximum
violation of CH inequalities. Now, providing that the state
vectors $\{ |S(\chi)\ra, |S(\chi+\pi) \ra ,|S(\chi')\ra,
|S(\chi'+\pi) \ra ,|S(\chi'')\ra, |S(\chi''+\pi) \ra , \dots \}$
($\chi=\alpha,\beta$) are chosen to be linearly independent, we
can extend the CH test to the $N$ pairs $\{(\alpha,
\beta),(\alpha', \beta'),(\alpha'', \beta''), \ldots
,(\alpha^{(N)}, \beta^{(N)})\}$ defining $N$ pairs of
two-dimensional subspaces whose union define a $2N\times 2N$
two-photon subspace.
In this way we can demonstrate the non-local nature of the
high-dimensional two-photon OAM-entangled states.

In summary, in this paper we proposed a novel experimental setup
to investigate the non-locality of high-dimensional  two-photon
OAM-entangled states generated by SPDC.
We use a pair of modified Mach-Zehnder
 interferometers (one per photon), as OAM analyzers.
 Inside each MZ there are two  SPPs (one per arm) which
 can rotate around their axes and  permit us to
explore the infinite-dimensional two-photon Hilbert
 space.
The output port of each MZ is made of a reflectivity-varying beam
splitter which acts as a polarizer in the two-dimensional space
defined by the two spatial modes (the two arms) of each MZ.
When the output ports of these OAM analyzers are fed into
single-mode optical fibers, the effective dimensionality of the
two-photon Hilbert space reduces from $\infty$ to $4$.
Because of this entanglement-preserving dimensional reduction, our
experimental scheme permits us to check the non-locality of the
two-photon OAM-entangled state, by using a $d \times N_a \times
N_b = 2 \times2 \times 2$ inequality \cite{Massar02a}.
In this way we found the maximum violation  of the CH inequality
for any four-dimensional two-photon subspace
 selected by the SPPs
orientations. Moreover, because of the strict analogy between ours
four-dimensional two-photon subspaces and  four-dimensional
two-photon \emph{polarization} space, other interesting
experiments (e.g. teleportation of spatial degrees of freedom) can
be implemented by using our scheme.
\begin{acknowledgments}
We acknowledge Richard Gill with whom we had insightful
discussions. We acknowledge support from the EU under the
IST-ATESIT contract. This project is also supported by FOM.
\end{acknowledgments}
\appendix
\section{}
For completeness, we give here explicit expressions for the
unnormalized probabilities displayed in Eq.
(\ref{eq380a}-\ref{eq380d}). Here $\delta = \alpha -\beta$.
\begin{equation}\label{a10}
\begin{array}{rcl}
\displaystyle{ P_{ab} } \displaystyle{(\theta_a, \theta_b)} &  =&
\displaystyle{\delta^2 \cos^2(\theta_a - \theta_b) -
2 \pi |\delta|\cos^2(\theta_a - \theta_b)} \\
&&  \displaystyle{  + \sin^2\theta_a \Bigl\{ \pi^2 \sin^2\theta_b
+ \cos^2\theta_b \Bigl[ 2 \pi^2 + \delta^2} \\
&& \displaystyle{  - 2 \pi \left( \delta + |\pi - \delta| \right)
\Bigr] \Bigr\} + \cos^2\theta_a \Bigl\{ \pi^2 \cos^2\theta_b
}\\
&&  \displaystyle{  + \sin^2\theta_b \Bigl[ 2 \pi^2 + \delta^2  -
2 \pi \left( -\delta + |\pi + \delta| \right) \Bigr] \Bigr\}}
\\
&& \displaystyle{  + \frac{1}{2} \sin (2 \theta_a) \sin
(2\theta_b) \Bigl[ \pi \left( |\pi + \delta| + |\pi - \delta|
\right)
}\\
&&  \displaystyle{  - |\pi + \delta| |\pi - \delta| \Bigr], }
  \end{array}
\end{equation}
%
\begin{equation}\label{a20}
\begin{array}{rcl}
\displaystyle{ P_{ab}(\theta_a, \infty)} & = & \displaystyle{ 3
\pi^2 + 2\delta^2 -\pi \Bigl( |\pi + \delta| + 2|\delta|+ |\pi -
\delta| \Bigr)  } \\
&&  \displaystyle{+ \pi \Bigl[ 2\delta -|\pi + \delta| + |\pi -
\delta| \Bigr] \cos (2 \theta_a) },
  \end{array}
\end{equation}
%
%
\begin{equation}\label{a30}
\begin{array}{rcl}
\displaystyle{ P_{ab}(\infty, \theta_b )} & = & \displaystyle{ 3
\pi^2 + 2\delta^2 -\pi \Bigl( |\pi + \delta| + 2|\delta|+ |\pi -
\delta| \Bigr)  } \\
&&  \displaystyle{+ \pi \Bigl[ 2\delta -|\pi + \delta| + |\pi -
\delta| \Bigr] \cos (2 \theta_b) },
  \end{array}
\end{equation}
%
%
\begin{equation}\label{a40}
 P_{ab}(\infty, \infty ) =   6 \pi^2
+ 4\delta^2 - 2 \pi \Bigl( |\pi + \delta| + 2|\delta|+ |\pi -
\delta| \Bigr).
\end{equation}
%
%
%

\begin{thebibliography}{28}
\expandafter\ifx\csname
natexlab\endcsname\relax\def\natexlab#1{#1}\fi
\expandafter\ifx\csname bibnamefont\endcsname\relax
  \def\bibnamefont#1{#1}\fi
\expandafter\ifx\csname bibfnamefont\endcsname\relax
  \def\bibfnamefont#1{#1}\fi
\expandafter\ifx\csname citenamefont\endcsname\relax
  \def\citenamefont#1{#1}\fi
\expandafter\ifx\csname url\endcsname\relax
  \def\url#1{\texttt{#1}}\fi
\expandafter\ifx\csname
urlprefix\endcsname\relax\def\urlprefix{URL }\fi
\providecommand{\bibinfo}[2]{#2}
\providecommand{\eprint}[2][]{\url{#2}}

\bibitem[{\citenamefont{Nielsen and Chuang}(2002)}]{NielsenBook}
\bibinfo{author}{\bibfnamefont{M.~A.} \bibnamefont{Nielsen}} \bibnamefont{and}
  \bibinfo{author}{\bibfnamefont{I.~L.} \bibnamefont{Chuang}},
  \emph{\bibinfo{title}{Quantum Computation and Quantum Information}}
  (\bibinfo{publisher}{Cambridge University Press},
  \bibinfo{address}{Cambridge, UK}, \bibinfo{year}{2002}),
  \bibinfo{edition}{reprinted first} ed.

\bibitem[{\citenamefont{Gisin et~al.}(2002)\citenamefont{Gisin, Ribody, Tittel,
  and Zbinden}}]{Gisin02}
\bibinfo{author}{\bibfnamefont{N.}~\bibnamefont{Gisin}},
  \bibinfo{author}{\bibfnamefont{G.}~\bibnamefont{Ribody}},
  \bibinfo{author}{\bibfnamefont{W.}~\bibnamefont{Tittel}}, \bibnamefont{and}
  \bibinfo{author}{\bibfnamefont{H.}~\bibnamefont{Zbinden}},
  \bibinfo{journal}{Rev. Mod. Phys.} \textbf{\bibinfo{volume}{74}},
  \bibinfo{pages}{145} (\bibinfo{year}{2002}).

\bibitem[{\citenamefont{Kwiat et~al.}(1995)\citenamefont{Kwiat, Mattle,
  Weinfurter, Zeilinger, Sergienko, and Shih}}]{Kwiat95}
\bibinfo{author}{\bibfnamefont{P.~G.} \bibnamefont{Kwiat}},
  \bibinfo{author}{\bibfnamefont{K.}~\bibnamefont{Mattle}},
  \bibinfo{author}{\bibfnamefont{H.}~\bibnamefont{Weinfurter}},
  \bibinfo{author}{\bibfnamefont{A.}~\bibnamefont{Zeilinger}},
  \bibinfo{author}{\bibfnamefont{A.~V.} \bibnamefont{Sergienko}},
  \bibnamefont{and} \bibinfo{author}{\bibfnamefont{Y.}~\bibnamefont{Shih}},
  \bibinfo{journal}{Phys. Rev. Lett.} \textbf{\bibinfo{volume}{75}},
  \bibinfo{pages}{4337} (\bibinfo{year}{1995}).

\bibitem[{\citenamefont{Mair et~al.}(2001)\citenamefont{Mair, Vaziri, Weihs,
  and Zeilinger}}]{Mair01a}
\bibinfo{author}{\bibfnamefont{A.}~\bibnamefont{Mair}},
  \bibinfo{author}{\bibfnamefont{A.}~\bibnamefont{Vaziri}},
  \bibinfo{author}{\bibfnamefont{G.}~\bibnamefont{Weihs}}, \bibnamefont{and}
  \bibinfo{author}{\bibfnamefont{A.}~\bibnamefont{Zeilinger}},
  \bibinfo{journal}{Nature (London)} \textbf{\bibinfo{volume}{412}},
  \bibinfo{pages}{313} (\bibinfo{year}{2001}).

\bibitem[{\citenamefont{Vaziri et~al.}(2002)\citenamefont{Vaziri, Weihs, and
  Zeilinger}}]{Vaziri02a}
\bibinfo{author}{\bibfnamefont{A.}~\bibnamefont{Vaziri}},
  \bibinfo{author}{\bibfnamefont{G.}~\bibnamefont{Weihs}}, \bibnamefont{and}
  \bibinfo{author}{\bibfnamefont{A.}~\bibnamefont{Zeilinger}},
  \bibinfo{journal}{Phys. Rev. Lett.} \textbf{\bibinfo{volume}{89}},
  \bibinfo{pages}{240401} (\bibinfo{year}{2002}).

\bibitem[{\citenamefont{de~Riedmatten et~al.}(2002)\citenamefont{de~Riedmatten,
  Marcikic, Zbinden, and Gisin}}]{Gisin02b}
\bibinfo{author}{\bibfnamefont{H.}~\bibnamefont{de~Riedmatten}},
  \bibinfo{author}{\bibfnamefont{I.}~\bibnamefont{Marcikic}},
  \bibinfo{author}{\bibfnamefont{H.}~\bibnamefont{Zbinden}}, \bibnamefont{and}
  \bibinfo{author}{\bibfnamefont{N.}~\bibnamefont{Gisin}},
  \bibinfo{journal}{Quantum Inf. Comput.} \textbf{\bibinfo{volume}{2}},
  \bibinfo{pages}{425} (\bibinfo{year}{2002}).

\bibitem[{\citenamefont{Law and Eberly}(2004)}]{Law04}
\bibinfo{author}{\bibfnamefont{C.~K.} \bibnamefont{Law}} \bibnamefont{and}
  \bibinfo{author}{\bibfnamefont{J.~H.} \bibnamefont{Eberly}},
  \bibinfo{journal}{Phys. Rev. Lett.} \textbf{\bibinfo{volume}{92}},
  \bibinfo{pages}{127903} (\bibinfo{year}{2004}).

\bibitem[{Pad()}]{Padua}
\bibinfo{note}{L. Neves, S. P\'{a}dua, and C. Saavedra, Phys. Rev. A
  \textbf{69}, 042305 (2004); L. Neves, G. Lima, J. G. Aguirre G\'{o}mez, C. H.
  Monken, C. Saavedra, and S. P\'{a}dua, Phys. Rev. Lett. \textbf{94}, 100501
  (2005).}

\bibitem[{\citenamefont{O'Sullivan-Hale
  et~al.}(2005)\citenamefont{O'Sullivan-Hale, Khan, Boyd, and
  Howell}}]{Boyd05a}
\bibinfo{author}{\bibfnamefont{M.~N.} \bibnamefont{O'Sullivan-Hale}},
  \bibinfo{author}{\bibfnamefont{I.~A.} \bibnamefont{Khan}},
  \bibinfo{author}{\bibfnamefont{R.~W.} \bibnamefont{Boyd}}, \bibnamefont{and}
  \bibinfo{author}{\bibfnamefont{J.~C.} \bibnamefont{Howell}},
  \bibinfo{journal}{Phys. Rev. Lett.} \textbf{\bibinfo{volume}{94}},
  \bibinfo{pages}{220501} (\bibinfo{year}{2005}).

\bibitem[{Aci()}]{Acin03a}
\bibinfo{note}{A. Ac\'{i}n, N. Gisin, L. Masanes, V. Scarani,
  arXiv:quant-ph/0310166 (2003).}

\bibitem[{\citenamefont{Oemrawsingh et~al.}(2004)\citenamefont{Oemrawsingh,
  Aiello, Eliel, Nienhuis, and Woerdman}}]{Sumant04b}
\bibinfo{author}{\bibfnamefont{S.~S.~R.} \bibnamefont{Oemrawsingh}},
  \bibinfo{author}{\bibfnamefont{A.}~\bibnamefont{Aiello}},
  \bibinfo{author}{\bibfnamefont{E.~R.} \bibnamefont{Eliel}},
  \bibinfo{author}{\bibfnamefont{G.}~\bibnamefont{Nienhuis}}, \bibnamefont{and}
  \bibinfo{author}{\bibfnamefont{J.~P.} \bibnamefont{Woerdman}},
  \bibinfo{journal}{Phys. Rev. Lett.} \textbf{\bibinfo{volume}{92}},
  \bibinfo{pages}{217901} (\bibinfo{year}{2004}).

\bibitem[{sum()}]{sumant05c}
\bibinfo{note}{S. S. R. Oemrawsingh, X. Ma, D. Voigt, A. Aiello, E. R. Eliel,
  G. W. 't Hooft, and J. P. Woerdman, arXiv:quant-ph/0506253.}

\bibitem[{\citenamefont{Clauser and Horne}(1974)}]{Clauser74a}
\bibinfo{author}{\bibfnamefont{J.}~\bibnamefont{Clauser}} \bibnamefont{and}
  \bibinfo{author}{\bibfnamefont{M.~A.} \bibnamefont{Horne}},
  \bibinfo{journal}{Phys. Rev. D} \textbf{\bibinfo{volume}{10}},
  \bibinfo{pages}{526} (\bibinfo{year}{1974}).

\bibitem[{\citenamefont{Jeffers et~al.}(1993)\citenamefont{Jeffers, Imoto, and
  Loudon}}]{Jeffers93a}
\bibinfo{author}{\bibfnamefont{J.~R.} \bibnamefont{Jeffers}},
  \bibinfo{author}{\bibfnamefont{N.}~\bibnamefont{Imoto}}, \bibnamefont{and}
  \bibinfo{author}{\bibfnamefont{R.}~\bibnamefont{Loudon}},
  \bibinfo{journal}{Phys. Rev. A} \textbf{\bibinfo{volume}{47}},
  \bibinfo{pages}{3346} (\bibinfo{year}{1993}).

\bibitem[{\citenamefont{Beijersbergen et~al.}(1994)\citenamefont{Beijersbergen,
  Coerwinkel, Kristensen, and Woerdman}}]{Beije94a}
\bibinfo{author}{\bibfnamefont{M.~W.} \bibnamefont{Beijersbergen}},
  \bibinfo{author}{\bibfnamefont{R.~P.~C.} \bibnamefont{Coerwinkel}},
  \bibinfo{author}{\bibfnamefont{M.}~\bibnamefont{Kristensen}},
  \bibnamefont{and} \bibinfo{author}{\bibfnamefont{J.~P.}
  \bibnamefont{Woerdman}}, \bibinfo{journal}{Opt. Commun.}
  \textbf{\bibinfo{volume}{112}}, \bibinfo{pages}{321} (\bibinfo{year}{1994}).

\bibitem[{\citenamefont{Campos et~al.}(1989)\citenamefont{Campos, Saleh, and
  Teich}}]{Campos89a}
\bibinfo{author}{\bibfnamefont{R.~A.} \bibnamefont{Campos}},
  \bibinfo{author}{\bibfnamefont{B.~E.~A.} \bibnamefont{Saleh}},
  \bibnamefont{and} \bibinfo{author}{\bibfnamefont{M.~C.} \bibnamefont{Teich}},
  \bibinfo{journal}{Phys. Rev. A} \textbf{\bibinfo{volume}{40}},
  \bibinfo{pages}{1371} (\bibinfo{year}{1989}).

\bibitem[{\citenamefont{Walborn et~al.}(2003)\citenamefont{Walborn,
  de~Oliveira, P\'{a}dua, and Monken}}]{Walborn03a}
\bibinfo{author}{\bibfnamefont{S.~P.} \bibnamefont{Walborn}},
  \bibinfo{author}{\bibfnamefont{A.~N.} \bibnamefont{de~Oliveira}},
  \bibinfo{author}{\bibfnamefont{S.}~\bibnamefont{P\'{a}dua}},
  \bibnamefont{and} \bibinfo{author}{\bibfnamefont{C.~H.}
  \bibnamefont{Monken}}, \bibinfo{journal}{Phys. Rev. Lett.}
  \textbf{\bibinfo{volume}{90}}, \bibinfo{pages}{143601}
  (\bibinfo{year}{2003}).

\bibitem[{Dan()}]{Dang05a}
\bibinfo{note}{Gui-F. Dang, Li-P. Deng, and Kaige Wang, arXiv:quanth-ph/0504057
  (2005)}.

\bibitem[{\citenamefont{Visser and Nienhuis}(2004)}]{Visser04a}
\bibinfo{author}{\bibfnamefont{J.}~\bibnamefont{Visser}} \bibnamefont{and}
  \bibinfo{author}{\bibfnamefont{G.}~\bibnamefont{Nienhuis}},
  \bibinfo{journal}{Eur. Phys. J. D} \textbf{\bibinfo{volume}{29}},
  \bibinfo{pages}{301} (\bibinfo{year}{2004}).

\bibitem[{\citenamefont{Kim et~al.}(2002)\citenamefont{Kim, Son, Bu\v{z}ek, and
  Knight}}]{Kim02a}
\bibinfo{author}{\bibfnamefont{M.~S.} \bibnamefont{Kim}},
  \bibinfo{author}{\bibfnamefont{W.}~\bibnamefont{Son}},
  \bibinfo{author}{\bibfnamefont{V.}~\bibnamefont{Bu\v{z}ek}},
  \bibnamefont{and} \bibinfo{author}{\bibfnamefont{P.~L.}
  \bibnamefont{Knight}}, \bibinfo{journal}{Phys. Rev. A}
  \textbf{\bibinfo{volume}{65}}, \bibinfo{pages}{032323}
  (\bibinfo{year}{2002}).

\bibitem[{\citenamefont{Bell}(1965)}]{Bell1}
\bibinfo{author}{\bibfnamefont{J.~S.} \bibnamefont{Bell}},
  \bibinfo{journal}{Physics (N.Y.)} \textbf{\bibinfo{volume}{1}},
  \bibinfo{pages}{195} (\bibinfo{year}{1965}).

\bibitem[{\citenamefont{Clauser et~al.}(1969)\citenamefont{Clauser, Horne,
  Shimony, and Holt}}]{Clauser69}
\bibinfo{author}{\bibfnamefont{J.~F.} \bibnamefont{Clauser}},
  \bibinfo{author}{\bibfnamefont{M.~A.} \bibnamefont{Horne}},
  \bibinfo{author}{\bibfnamefont{A.}~\bibnamefont{Shimony}}, \bibnamefont{and}
  \bibinfo{author}{\bibfnamefont{R.~A.} \bibnamefont{Holt}},
  \bibinfo{journal}{Phys. Rev. Lett.} \textbf{\bibinfo{volume}{23}},
  \bibinfo{pages}{880} (\bibinfo{year}{1969}).

\bibitem[{\citenamefont{Garuccio and Rapisarda}(1981)}]{GaruccioeRapisarda}
\bibinfo{author}{\bibfnamefont{A.}~\bibnamefont{Garuccio}} \bibnamefont{and}
  \bibinfo{author}{\bibfnamefont{V.~A.} \bibnamefont{Rapisarda}},
  \bibinfo{journal}{Il Nuovo Cimento} \textbf{\bibinfo{volume}{65A}},
  \bibinfo{pages}{269} (\bibinfo{year}{1981}).

\bibitem[{Aie()}]{Aiello05d}
\bibinfo{note}{A. Aiello, S. S. R. Oemrawsingh, E. R. Eliel, and J. P.
  Woerdman, arXiv:quant-ph/0503034 (2005)}.

\bibitem[{Sum()}]{Sumant04a}
\bibinfo{note}{S. S. R. Oemrawsingh, A. Aiello, E. R. Eliel, and J. P.
  Woerdman, arXiv:quant-ph/0401148 (2004).}

\bibitem[{\citenamefont{\.{Z}ukowski et~al.}(1997)\citenamefont{\.{Z}ukowski,
  Zeilinger, and Horne}}]{Zuko97a}
\bibinfo{author}{\bibfnamefont{M.}~\bibnamefont{\.{Z}ukowski}},
  \bibinfo{author}{\bibfnamefont{A.}~\bibnamefont{Zeilinger}},
  \bibnamefont{and} \bibinfo{author}{\bibfnamefont{M.~A.} \bibnamefont{Horne}},
  \bibinfo{journal}{Phys. Rev. A} \textbf{\bibinfo{volume}{55}},
  \bibinfo{pages}{2564} (\bibinfo{year}{1997}).

\bibitem[{\citenamefont{Peres}(1978)}]{Peres78a}
\bibinfo{author}{\bibfnamefont{A.}~\bibnamefont{Peres}}, \bibinfo{journal}{Am.
  J. Phys.} \textbf{\bibinfo{volume}{46}}, \bibinfo{pages}{745}
  (\bibinfo{year}{1978}).

\bibitem[{\citenamefont{Massar et~al.}(2002)\citenamefont{Massar, Pironio,
  Roland, and Gisin}}]{Massar02a}
\bibinfo{author}{\bibfnamefont{S.}~\bibnamefont{Massar}},
  \bibinfo{author}{\bibfnamefont{S.}~\bibnamefont{Pironio}},
  \bibinfo{author}{\bibfnamefont{J.}~\bibnamefont{Roland}}, \bibnamefont{and}
  \bibinfo{author}{\bibfnamefont{B.}~\bibnamefont{Gisin}},
  \bibinfo{journal}{Phys. Rev. A} \textbf{\bibinfo{volume}{66}},
  \bibinfo{pages}{052112} (\bibinfo{year}{2002}).

\end{thebibliography}
%

\end{document}